\documentclass[11pt,fleqn]{article}

\usepackage{amssymb}
\def\noi{\noindent}

\newcommand{\Title}[1]{\noi {\uppercase{\Large #1}}\\[1ex]}

\newcommand{\Author}[2]{\noi{\large\bf #1}\\[2ex]\noi{\normalsize\it #2}\\}

\newcommand{\Abstract}[1]{\vskip 2mm \begin{center}
        \parbox{16.4cm}{\small\noi #1} \end{center}\medskip}
\newcommand{\PACS}[1]{\begin{center}{\small PACS: #1}\end{center}}
\newcommand{\foom}[1]{\protect\footnotemark[#1]}
\newcommand{\foox}[2]{\footnotetext[#1]{#2}\addtocounter{footnote}{1}}
\def\email#1#2{\footnotetext[#1]{e-mail: #2}\addtocounter{footnote}{1}}

\def\Talk{\foox 1 {Talk given at the International Conference RUSGRAV-13,
	   June 23--28, 2008, PFUR, Moscow}}

\def\nqq{\hspace*{-2em}}

\def\Jl#1#2{#1 {\bf #2},\ }

\def\ApJ#1 {\Jl{Astroph. J.}{#1}}
\def\CQG#1 {\Jl{Class. Quantum Grav.}{#1}}
\def\DAN#1 {\Jl{Dokl. AN SSSR}{#1}}
\def\GC#1 {\Jl{Grav. Cosmol.}{#1}}
\def\GRG#1 {\Jl{Gen. Rel. Grav.}{#1}}
\def\JETF#1 {\Jl{Zh. Eksp. Teor. Fiz.}{#1}}
\def\JETP#1 {\Jl{Sov. Phys. JETP}{#1}}
\def\JHEP#1 {\Jl{JHEP}{#1}}
\def\JMP#1 {\Jl{J. Math. Phys.}{#1}}
\def\NPB#1 {\Jl{Nucl. Phys. B}{#1}}
\def\NP#1 {\Jl{Nucl. Phys.}{#1}}
\def\PLA#1 {\Jl{Phys. Lett. A}{#1}}
\def\PLB#1 {\Jl{Phys. Lett. B}{#1}}
\def\PRD#1 {\Jl{Phys. Rev. D}{#1}}
\def\PRL#1 {\Jl{Phys. Rev. Lett.}{#1}}

\def\lal{&&\nqq {}}
\def\eq{Eq.\,}

\def\beq{\begin{equation}}
\def\eeq{\end{equation}}
\def\bear{\begin{eqnarray}}
\def\bearr{\begin{eqnarray} \lal}
\def\ear{\end{eqnarray}}
\def\earn{\nonumber \end{eqnarray}}

\def\nnn{\nonumber\\ \lal }

\def\yy{\\[5pt] {}}

\def\d{\partial}

\def\eps{\varepsilon}

\begin{document}
\twocolumn[

\Title{On unification of gravitation and electromagnetism\yy
	in the framework of a general-relativistic approach\foom 1}

\Author{Alexander A. Chernitskii\foom 2}
{Friedmann Laboratory of Theoretical Physics, St. Petersburg;\\
 State University of Engineering and Economics}

%\Rec{October 10, 2008}

\Abstract
{We consider the unification problem for the gravitational and
 electromagnetic interactions and its possible solution on the basis of the
 existence of an effective Riemannian space in nonlinear electrodynamics

 }

] %%%%%%%%%%%%%%%%%%%%%%%%%%
\Talk
\email 2 {AAChernitskii@mail.ru,\\ AAChernitskii@engec.ru}

\section{Introduction}

  The problem of unification of all interactions in a unified theory is
  remaining to be one of the most important problems of theoretical physics.
  Such a unification in a theory of some unified field seems to deserve a
  serious attention.

  Among the interactions on which we now can speak, it is natural to single
  out two long-range interactions, gravitation and electromagnetism.
  Obviously, in a unified field theory these two interactions should appear
  in a natural way.

  The circumstance that the gravitational interaction is related to changes
  in the symmetric metric tensor of space-time is nowadays a generally
  accepted fact. It must be noted, however, that for an agreement with the
  experimental data it is sufficient to assume that, for the motion of
  material bodies, there is an effective Riemannian space which can be, in
  turn, induced by a certain field different from the metric tensor field.

  Such an induced Riemannian space appears in nonlinear electrodynamics when
  one considers the long-range interaction of particles which are solitons of
  the model [1, 2].

  According to different viewpoints, the electrodynamics of vacuum should be
  regarded nonlinear. This means that the above-mentioned effective
  Riemannian space must be at least taken into account in calculations
  related to the gravitational interaction. But a radical viewpoint will be
  that gravitation as a whole is a manifestation of the electromagnetic field
  nonlinearity.

\section{A generally invariant nonlinear field model}

  Consider the field of a certain second-rank tensor $G_{\mu\nu}$ in a
  four-dimensional space-time and the following variational principle:
\beq                                      \label{chern-43217400}
	\delta\int\!\!\sqrt{|\det(G_{\mu\nu})|}\;(\mathrm{d}x)^{4} = 0,
\eeq
  where $(\mathrm{d}x)^{4}\equiv \mathrm{d}x^0\mathrm{d}x^1\mathrm{d}
  x^2\mathrm{d}x^3$, and Greek indices take the values $0,\ 1,\ 2,\ 3$.

  Due to the rule of variable changing in an integral and the determinant
  transformation rule, this variational principle is invariant with respect
  to general coordinate transformations.

  The tensor $G_{\mu\nu}$ may be represented as a sum of symmetric and
  antisymmetric tensors, of which the first one could be identified with
  the metric tensor and the second one with the electromagnetic field tensor:
\beq                                           \label{chern-44724464}
	G_{\mu\nu} = g_{\mu\nu} + \chi^2\,F_{\mu\nu},
	\qquad
  F_{\mu\nu} = \frac{\d A_{\nu}}{\d x^{\mu}} - \frac{\d A_{\mu}}{\d x^{\nu}},
\eeq
  $A_{\mu}$ being the components of the electromagnetic potential.

  The variational principle of the form (1) was considered by A.S. Eddington
  [3] and A. Einstein [4] just because of its general invariance.

  M. Born and L. Infeld [5] considered the variational principle with the
  tensor $G_{\mu\nu}$ of the form (2), where $g_{\mu\nu}$ is the metric
  tensor of flat space.

\section{Equations of nonlinear electrodynamics and the effective Riemannian
	space}

  The Born-Infeld set of equations has the form
\beq                                           \label{chern-35357121}
	\frac{\d}{\d x^\mu}\,\sqrt{|g|}\; f^{\mu\nu} {}={} 0,
\eeq
  where
\bearr                                                  \label{chern-65713937}
	f^{\mu\nu} \equiv \frac{\chi^{-2}\,\d{\cal L}}{\d(\d_\mu A_\nu)}
	= \frac{1}{\cal L}\,\left(F^{\mu\nu} -
  \frac{\chi^2}{2}\,\cal J\,\eps^{\mu\nu\sigma\rho}\,F_{\sigma\rho}\right),
\nnn
	{\cal L} \equiv  \sqrt{|\,1  -\chi^2\,{\cal I}
				     -\chi^4\,{\cal J}^2\,|},
\nnn
     {\cal I}  \equiv F_{\mu\nu}\,F^{\nu\mu}/2,\ \ \
     {\cal J}  \equiv \eps_{\mu\nu\sigma\rho}\, F^{\mu\nu}F^{\sigma\rho}/8,
\nnn
     \eps_{0123} \equiv \sqrt{|g|}, \ \ \ \eps^{0123}  = - 1/\sqrt{|g|}.
\earn

  The Born-Infeld model has the following energy-momentum tensor:
\beq                                          \label{chern-71416389}
	T^\mu_{.\nu} \equiv  \left(f^{\mu\rho}\,F_{\nu\rho}   -
	\chi^{-2}\,\left({\cal L}  -  1\right)\,\delta^\mu_\nu\right)/4\pi.
\eeq

  Remarkable is the characteristic equation in Born-Infeld electrodynamics
  [6]:
\beq                                             \label{chern-CharEq}
  \tilde{g}^{\mu\nu}\,\frac{\d \Phi}{\d x^\mu}\,\frac{\d \Phi}{\d x^\nu}=0,
\eeq
   where $\Phi (x^\mu)=0$ is the equation of the characteristic surface and
\beq                                         \label{chern-73144072}
	\tilde{g}^{\mu\nu} \equiv g^{\mu\nu} - 4\pi\,\chi^2\,T^{\mu\nu}.
\eeq
  Here, $T^{\mu\nu}$ is the energy-momentum tensor of the form (5).

\section{On unification of gravitation and electromagnetism}

  In connection with the form of \eq (5), the symmetric tensor
  $\tilde{g}^{\mu\nu}$ can be named the metric of the effective Riemannian
  space. Indeed, in studying a multiparticle solution of the model with
  the perturbation method, it turns out that the rapidly oscillating soliton
  particle propagates along geodesics of the effective Riemannian space with
  the metric $\tilde{g}^{\mu\nu}$ induced by remote particles [2]. Rays of
  high-frequency electromagnetic waves are bended in the same way as in a
  gravitational field with the metric $\tilde{g}^{\mu\nu}$.

  These results mean that in nonlinear electrodynamics there is an
  interaction which is indistinguishable from the gravitational one.

  Since the components of the energy-momentum tensor depend on even powers
  of the electromagnetic field tensor components, this interaction appears
  in the second order with respect to the weak field of remote solitons.

  As to the electromagnetic interaction between soliton particles in this
  model, it naturally appears in the first order with respect to the weak
  field of remote solitons [2].

  Thus there are two long-range interactions in this nonlinear electrodynamic
  model, and they appear in the first and second orders in the weak field of
  remote solitons. These interactions may be identified as the
  electromagnetic and gravitational ones. And that is how these two
  well-known long-range interactions are unified.

  As already indicated, this viewpoint is radical in this approach.

  A more moderate viewpoint is that the effective metric gives only a part of
  the gravitational interaction.

  Adhering to this viewpoint, it is necessary to advert to the generally
  invariant principle (1) in the context of A. Einstein's article [4]. This
  article uses the second-rank tensor $R_{\mu\nu}$ built from the connections
  $\Gamma^{\alpha}_{\beta\gamma}$ which are the field functions of the
  theory. The tensor $R_{\mu\nu}$ is not symmetric. Further on, the tensor
  $R_{\mu\nu}$ is split into symmetric and antisymmetric parts. The symmetric
  part is identified with the metric and the antisymmetric one with the
  electromagnetic field. Variation with respect to
  $\Gamma^{\alpha}_{\beta\gamma}$ leads to the gravitational field equations
  and the Maxwell equations for the case of weak fields.

  It is evident that the electrodynamic equation of this theory are nonlinear
  for fields which are not weak. It is also evident that the nonlinear
  electromagnetic field will create an effective metric which will contribute
  to the gravitational interaction.

\section{On correspondence with experimental data}

  The effective metric $\tilde{g}^{\mu\nu}$ can give a gravitational
  potential which behaves as $(1/r)$ at infinity [7]. This is possible by
  taking into account the rapidly oscillating electromagnetic background
  and by averaging over the rapid oscillations [8]. In this case, the
  gravitational constant depends on the magnitude of the radiation
  background.

  It is also necessary here to touch upon gravitational waves which are
  intensively sought for. An indirect proof of their existence is, by common
  views, the observed energy loss by a gravitationally bound binary system.

  However, in the framework of the unified theory under consideration,
  the energy loss by such a system does not necessarily mean that it is
  transferred to gravitational waves. With any of the two viewpoints (radical
  or moderate in the above-mentioned sense), the lost energy of a binary
  system may pass (partly or completely) into the electromagnetic background.

  It is well known that gravitational waves have not yet been observed
  directly. The absence of their direct detection may be an argument in favor
  of the viewpoint that gravity is entirely created by the electromagnetic
  field.

\section{Conclusion}

  The above-described approach to the important problem of unification of
  interactions looks reasonable and requires further development. A
  comparison of new theoretical results with experimental data will enable us
  to judge upon the applicability of the theory.

\end{document}